\documentstyle[psbox, epsfig]{WORKSHOP}

\begin{document}

\title{
MAXI, LOFAR and Microquasars\\ 
{\large\sf  -- All-sky monitoring of X-ray binaries in X-rays and radio --} 
}

\author{
Rob Fender$^{1,2}$\\
\\[12pt]  
%
$^1$  School of Physics \& Astronomy, University of Southampton, UK\\
$^2$  Astronomical Institute `Anton Pannekoek', University of Amsterdam, NL\\
%
{\it E-mail: r.fender@soton.ac.uk} 
}

\abst{ In this paper I will discuss future synergies between the
  Japanese MAXI X-ray all-sky monitor, to be placed on the
  International Space Station in 2009, and the next-generation radio
  astronomy array LOFAR, currenlty under construction in The
  Netherlands. The wide fields of view and multiple beams of LOFAR
  will allow, in combination with MAXI, simultaneous all-sky
  monitoring of the sky in both the X-ray and radio bands. Focussing
  on microquasars, X-ray binary jet systems, I discuss how the
  combination of MAXI and LOFAR will help us to further understand the
  accretion--outflow coupling in such systems.
}

\kword{MAXI -- LOFAR -- Microquasars}

\maketitle
\thispagestyle{empty}

\section{Introduction}

Wide-field and/or all-sky monitoring in X-ray/$\gamma$-rayss, in
particular (but not exclusively) with CGRO BATSE and RXTE ASM, has
produced a vast array of new, valuable, and in many cases unexpected,
results. MAXI will carry forward this approach in X-ray astronomy,
with all-sky monitoring in the soft X-ray band for at least three
years from 2009.

Beyond the X-ray/$\gamma$-ray regime, all-sky monitoring has proved
much harder to achieve, primarily due to the small fields of view
achievable with focussing instruments at optical, infared and radio
wavelengths. However, the new generation of radio telescopes will in
large part be optimized for such wide-field surveying and monitoring,
and will provide for the first time something close to all-sky
monitoring at radio wavelengths (unfortunately the planet always
blocks some of the sky to these earthbound facilities). In particular,
the LOFAR {\em Radio Sky Monitor} is proposed to provide $\sim$daily
monitoring of over 60\% of the sky at the $\sim$mJy level.

Why do we care about complementary X-ray and radio observations ?
Nearly all phenomena associated with flaring and transient X-ray
behaviour are in fact associated with radio emission, and each
wavelength regime offers a different but complementary diagnostic of
the processes behind the transient outburst. In many cases, such as
accreting systems, the transient X-ray behaviour is associated with an
increase in the instantaneous accretion rate onto the central compact
object, and the radio emission is synchrotron emission resulting from
particle acceleration and outflow in a jet. Comparing the two allows
us to make quantitative estimates of the accretion power being
liberated in radiation, kinetic outflows or, in the case of black
holes, possibly being advected across an event horizon
(e.g. K\"ording, Fender \& Migliari 2006). In addition to this
incoherent synchrotron emission, coherent emission with shorter
durations and much higher brightness temperatures is associated with
e.g. pulsars and may have transient counterparts in other high-energy
sources which have not yet been detected (for example the nature of
the coherent extragalactic radio burst reported by Lorimer et
al. [2007] remains unclear).

\subsection{Microquasars}

The term `microquasar' has been profitably adopted for X-ray binary
systems with jets since Mirabel et al. (1992). Since all classes of
X-ray binary, with the exception of high-field X-ray pulsars, appear
to show jets (e.g. Fender 2006) the terms `microquasar' and `X-ray
binary' are almost synonymous. These systems are extreme examples of
the accretion:outflow scenario discussed above, and often undergo
outbursts in which a $\sim 10 M_{\odot}$ black hole brightens to close
to its Eddington limit ($\sim 10^{39}$ erg s$^{-1}$ whilst producing,
during certain phases of the outburst, a sporadic but very powerful
jet. As the name suggests, they may be compared quantitatively with
supermassive black holes in active galactic nuclei (AGN), allowing us
to understand the (rather straightforward) mass scaling relations in
black hole accretion (Merloni, Heinz \& di Matteo 2003; K\"ording,
Falcke \& Corbel 2006; McHardy et al. 2007). In addition,
understanding such scaling may allow us to use insights learned from
the binaries to understand the cosmological evolution of black hole
accretion and feedback.

In the past decade a rough phenomenological understanding of the
relation between X-rays and radio emission in such systems has
developed, summarized most recently in Fender, Belloni \& Gallo
(2004). Below about $\sim 1$\% of the Eddington luminosity, systems
exist in a `hard' X-ray state in which relatively steady radio
emission indicates the presence of a long-lived, powerful
outflow. Above this luminosity, systems often enter a hysteretic track
involving a soft state with little core radio emission, and a powerful
relativistic ejection event during the transition from hard to soft
states (see e.g. Fig 1). Subsequently systems return to the `hard'
state track (in reality the picture is more complex and detailed than
sketched out here, of course; see e.g. Homan \& Belloni 2007 for more
examples of such tracks).

\section{MAXI}

As part of the proceedings of the MAXI workshop, the {\em Monitor of
  All-Sky X-ray Image} (e.g. Kawai et al. 1999; Matsuoka et al. 2007)
requires little introduction. MAXI will attach to the {\em Kibo}
exposed facility onboard the International Space Station (ISS) in
2009, scanning the entire sky once per ISS orbit and achieving
milliCrab sensitivities on a timescale of a week. The monitoring will
be made with both position sensitive gas-proportional counters for
2-30 keV X-rays and CCD cameras for 0.5-10 keV X-rays. Obvious targets
include X-ray binaries, AGN and other outbursting high-energy objects
such as soft gamma-ray repeaters, supernovae etc.

\begin{figure}
\centerline{\epsfig{file=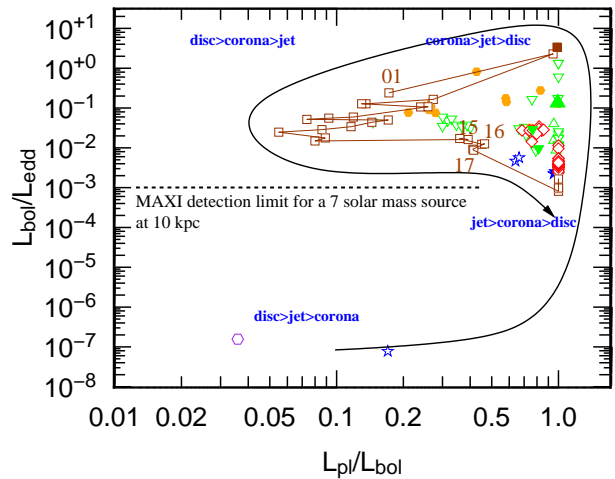, width=8cm}}
\caption{A `disc-fraction luminosity diagram' (DFLD; K\"ording Jester
\& Fender 2006) for a selection of X-ray binary systems (from Cabanac
et al. 2008). A naive description of the division of accretion power
between jet (kinetic), corona and accretion disc is indicated. Much of
the most interesting behaviour occurs above $\sim 10^{-3}$ Eddington
luminosity where the sources exhibit hysteresis and raapid switching
of spectral and jet modes. MAXI will be able to detect a typical X-ray
binary micoquasar source throughout the whole of this zone, with broad
(0.5--30 keV) spectral coverage. Combination of this monitoring with
radio observations will vastly improve our understanding of the
disc-jet connection in microquasars. In addition, for more nearby
sources MAXI will track the rise and decay to $\sim 10^{-4}$
Eddington, which is a very poorly sampled region of paramter space.}
\label{xmaxi}
\end{figure}

Much of the most interesting behaviour of X-ray binaries, and in
particular the disc-jet coupling, occurs in the zone of hysteresis
about $\sim 10^{-3}$ Eddington. MAXI will be able to track essentially
all galactic transients throughout this zone and onto the hard state
branch, providing a fantastic resource for our understanding of black
hole accretion (see Fig \ref{xmaxi}). Combination of the MAXI results
with those from radio monitoring programs, whether LOFAR or some other
facility, should provide new insights into, and refinements of, our
models for the accretion-outflow connection in relativistic objects.

\section{LOFAR}

\begin{figure}
\centerline{\epsfig{file=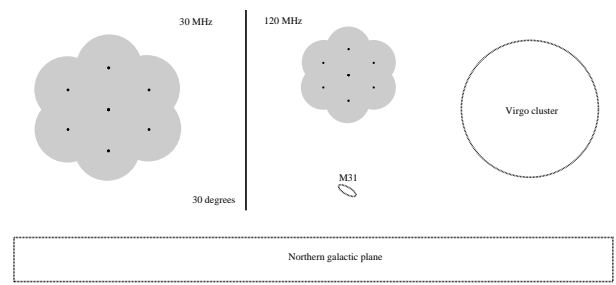, width=8cm}}
\caption{The vast fields of view achievable with LOFAR due to its low
frequency and multi-beam capability, compared to other astronomical
facilities and astrophysical objects of interest in the local
universe.}
\label{corebeam2}
\end{figure}

LOFAR is a next-generation radio telescope under construction in The
Netherlands with long-baseline stations under development in other
European countries (currently Germany, The UK, France, Sweden). The
array will operate in the 30--80 and 120--240 MHz bands (80--120 MHz
being dominated by FM radio transmissions in northern Europe). The
telescope is the flagship project for ASTRON, and is the largest of
the pathfinders for the lowest-frequency component of the Square
Kilometere Array (SKA). Core Station One (CS1; see Gunst et al. 2006)
is currently operating, and the next stage of deployment is about to
begin, with 36 stations to be in the field by the end of 2009.

LOFAR has an enormous field of view compared to previous radio
astronomy facilities (see Fig \ref{corebeam2}), allowing for the first
time wide-field repetitive monitoring of the radio sky, in the {\em
Radio Sky Monitor} mode (Fender et al. 2008).

For more information on the project, see:

\smallskip
{\bf http://www.lofar.org}
\smallskip

\section{Microquasars with MAXI and LOFAR}

The sensivitiy of MAXI will be about one order of magnitude better
than that of the RXTE ASM, and over a broader energy range,
representing a significant improvement. Many X-ray transients will be
detected and monitored by MAXI and, crucially, for nearby ($d \leq$
few kpc) sources MAXI will detect them while still below the $\sim
1$\% luminosity limit above which the complex hysteretical behaviour
occurs, thereby providing both early warning of outbursts and tracking
of the decay phase of systems.

\begin{figure}
\centerline{\epsfig{file=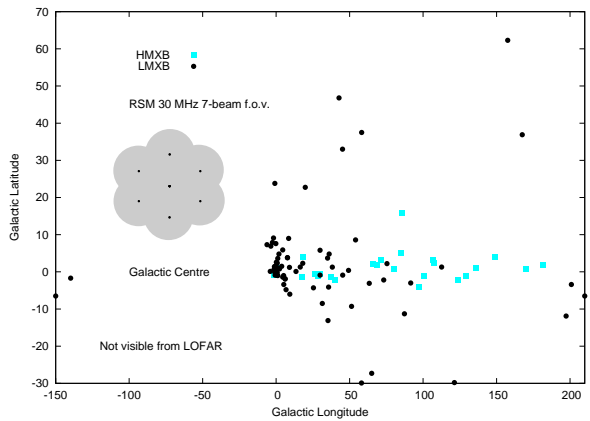, width=8cm}}
\caption{A comparison of the galactic plane visible from LOFAR, the
locations of known high-mass and low-mass X-ray binaries in the
visible region, and the seven-beam 30 MHZ field of view. LOFAR
pointings in the galactic plane will simultaneously measure many systems.}
\label{xrbbeam}
\end{figure}

The population of X-ray binaries accessible to LOFAR is limited by its
physical location in northern Europe, precluding observations of the
galactic centre region which contains many bright X-ray
binaries. Nevertheless, more than half of the galactic plane is
visible from LOFAR, containing many known X-ray binary systems (see
Fig \ref{xrbbeam}). 

\begin{figure}
\centerline{\epsfig{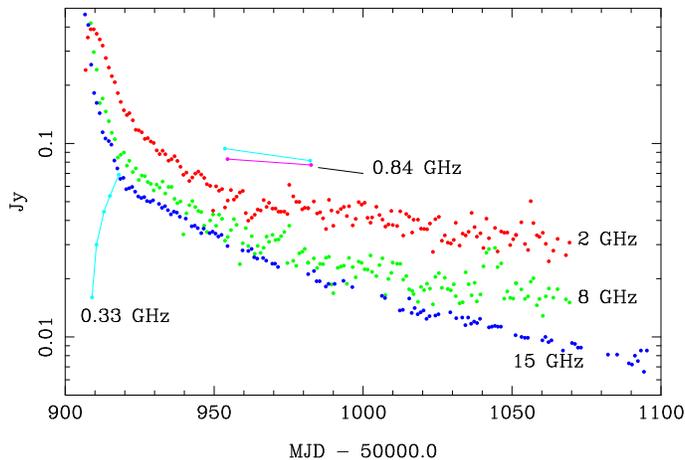}}
\caption{Radio observations of the ouburst of the X-ray transient CI
Cam, between 0.33 and 15 GHz. The lower frequencies, approaching the
LOFAR frequency range, peak much later than the higher frequencies due
to initial synchrotron self-absorption within the ejecta. This will
result in a delay between the major outburst/ejection events and their
detectability with LOFAR.}
\label{cicam}
\end{figure}

While the low observing frequencies of LOFAR facilitate the extremely
wide fields of view, they are not however optimum for observing X-ray
binaries or other sources of synchrotron-emitting ejecta. The reason
for this is that such ejecta are usually initially self-absorbed at
such low frequencies, and may take some significant time to peak in
the LOFAR band. In Fig \ref{cicam} we present the light curve of an
outburst from the binary system CI Cam (note: while the system itself
is unusual, the outburst is fairly typical for a synchrotron `bubble'
event). The flux density at 330 MHz, just above the LOFAR high band,
did not peak (corresponding to a transition from opticall thick to
optically thin) until several tens of days after the initial
outburst. Nevertheless, even the first observation at this frequency
was above the detection limit of the LOFAR RSM.

If we use MAXI to trigger LOFAR on the rise of a new, nearby,
transient, we should be able to track the flat-spectrum radio emission
from the system over two to three orders of magnitude at the highest
LOFAR frequencies (240 MHz; observations indicate the flat-spectrum
emission from Cyg X-1 extends between 0.33--220 GHz so this is not too
much of an extrapolation...). This will allow us to probe the
luminosity range around which several anomalously radio-faint systems
have been found (Gallo 2007) and so to further test and refine the
'universal' correlation in this state (Gallo, Fender \& Pooley
2003). Note that post-outburst, this correlation may well be
impossible to test as the emission in the LOFAR band may be dominated
by emission from ejecta launched earlier during state transitions (see
Fender, Belloni \& Gallo 2004). Therefore, good coordination between
MAXI and LOFAR is highly desirable in order to optimize their combined
science in this area.

\begin{figure}
\centerline{\epsfig{file=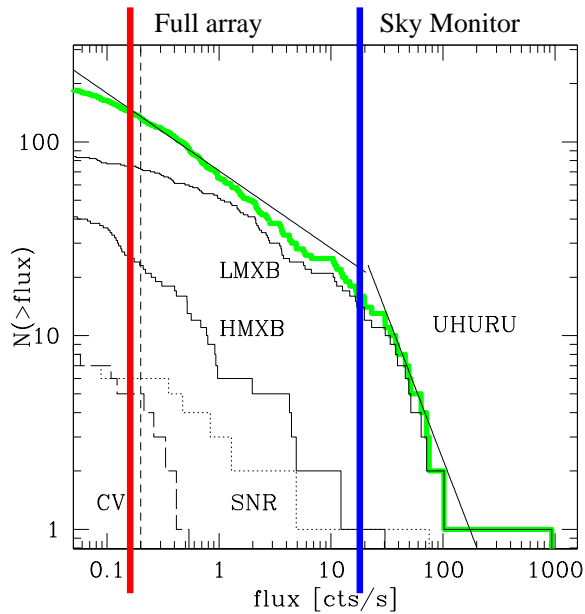, width=8cm}}
\caption{The cumulative distribution function of X-ray sources
detected by the RXTE ASM (Grimm, Gilfanov \& Sunyaev 2002), dominated
by LMXBs which are also likely to be jet (and therefore radio) sources
in outburst. MAXI will be about ten times more sensitive than RXTE
ASM, as a result of which more than 100 X-ray binaries should be
detectable per orbit of the ISS. The vertical dashed line, originally
placed to indicate the approximate completeness sample of the RXTE ASM
data, also corresponds approximately to the one-second sensivity limit
for MAXI. The red and blue vertical lines indicate an estimate of how
far into the population LOFAR can observe, based upon empirical
scalings between radio and X-ray luminosities in X-ray binaries. The
RSM mode should be able to detect the brightest $\sim 20$ sources in
each scan -- this will include semi-persistent systems such as Cyg
X-3, SS 433, as well the brightest binaries in outburst at that
epoch. Deeper observations with the full array will allow detailed
study of the entire binary population visible to MAXI, allowing
tracking of radio behaviour over at least three orders of magnitude in
luminosity for many sources.}
\label{lofarsens}
\end{figure}

How many sources might we expect ? Grimm, Gilfanov \& Sunyaev (2002)
derived the luminosity function for galactic X-ray binaries based upon
RXTE ASM data. This is replotted in Fig \ref{lofarsens} with
approximate sensitivity limits for different LOFAR modes. Based on
this it would be possible to monitor daily several hundred sources
with a combination of MAXI and both {\em Radio Sky Monitor} mode and
targetted observations of LOFAR. Even monitoring a few 10s of sources
({\em Radio Sky Monitor} mode alone) would be a very significant
resource indeed.

Following the major jet ejection, associated with the hard to soft
state transition (Fender, Belloni \& Gallo 2004), has occurred, LOFAR
is likely to be insensitive to the core once the ejecta have become
optically thin and completely dominate at low frequencies. This will
curtail attempts to track the core behaviour (for example, how much is
the core emission really suppressed in the soft state, and when does
it reactive in the soft to hard transition), which will be better done
with higher-frequency facilities (e.g. EVLA, e-MERLIN, ATA over the
MAXI lifetime). However, the emission at these frequencies is also
very long-lived, and may be the best way to look at the large-scale
jets which can develop over timescales of years and angular scales of
arcminutes (e.g. Corbel et al. 2002). With international baselines of
$\geq 1000$km, at 240 MHz LOFAR will have an angular resolution of
better than one arsec, plus very good sensitivity to low surface
brightness emission, and so will be an excellent instrument to observe
these jets as they decelerate and energise the ISM.

\section{Summary}

As a general tool for high-energy astrophysics, combined MAXI and
LOFAR monitoring of the (northern) sky in the X-ray and radio bands
will be etxremely valuable. Significant insights will be gained into
the physics behind numerous phenomena not touched on here, such as
soft gamma-ray repeaters, supernova, blazars etc. 

In the field of microquasars, the combination of MAXI and LOFAR will
probably be most apparent in two key areas:

\begin{enumerate}
\item{{\bf The radio:X-ray correlation in the hard state:} MAXI will
  detect a black hole X-ray transient at a distance of $\sim 10$ kpc
  to an Eddington ratio X-ray luminosity of about $10^{-3}$. During
  the rising phase of the outburst, assuming that the flat spectrum
  radio emission associated with the hard state extends to the LOFAR
  band ($\leq 240$ MHz) then we will gain a valuable new sample of
  data for the radio:X-ray correlation in hard states. Once the radio
  spectrum becomes dominated by ejecta (as occurs following the
  brightest canonical hard state), LOFAR observations are likely to be
  less useful for this correlation as synchrotron-emitting ejecta will
  evolve slowly at these frequencies, masking the level of core radio
  emission.}
\item{{\bf Spatially resolving extended jets:} Once the major ejection
  during the hard $\rightarrow$ soft transition occurs, ejecta from
  black hole binaries typically separate from the core with proper
  motions of 10--30 mas d$^{-1}$. With international baselines, at the
  highest frequency (240 MHz), LOFAR has an angular resolution of
  $\leq 1$ arsec, and so such eject should be resolvable after a
  couple of months. From this point onwards, with its sensitivity to
  optically thin radio emission and low surface brightness features,
  LOFAR may be the premium facility for emission the large-scale
  evolution of such ejecta as it interacts with the ISM (note that
  expanding radio emission on much large scales is still visible ten
  years after the 1998 outburst of XTE J1550-564: S. Corbel, private
  communication).}
\end{enumerate}

LOFAR is however much better suited to finding prompt {\em coherent}
radio emission, such as that associated with pulsars, flare stars
etc. and speculated to be associated with gamma-ray bursts,
relativistic object mergers etc. Very naively, were such bursts to be
associated with X-ray binary outbursts, we might expect them to occur
around the most violently variable phase, i.e. the hard $\rightarrow$
soft state transition. Therefore we add another, much more
speculative, point:

\begin{enumerate}
\setcounter{enumi}{2}
\item{{\bf Coherent bursts during state transitions:} there is a small
  chance that the violent processes associated with the state
  transitions, ejections, and reformation of the accretion flow, may
  be associated with coherent radio bursts.}
\end{enumerate}

\begin{figure}
\centerline{\epsfig{file=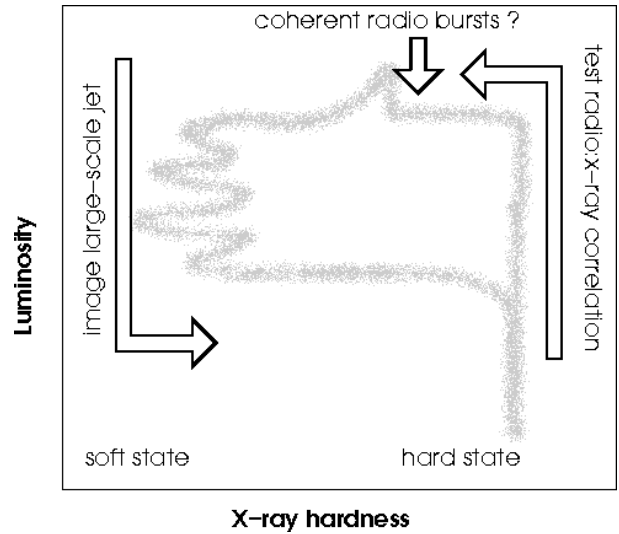, width=8cm}}
\caption{Schematic illustrating two areas where MAXI and LOFAR will
  impact upon studies of microquasar systems. MAXI should detect many
  black hole transients in the rising phase of the outburst, above
  about $\sim 0.1$\% Eddington. During this phase, until the
  transition to the soft state begins (left to right transition in the
  figure, typically between 1--30\% Eddington; see also Fig 1),
  combined MAXI and LOFAR observations will be able to test the hard
  state radio:X-ray correlation. Once radio flaring occurs, during the
  hard to soft transition, the low-frequency LOFAR radio bands will be
  dominated for a long period of time by slowly evolving synchrotron
  emission and will not be useful for studying the
  correlation. However, after a couple of months any large-scale
  ejecta will be resolvable using the international LOFAR baselines,
  allowing us to trace the evolution of deceleration and in-situ
  particle acceleration as the jets interact with the ISM. Much more
  speculatively, during the explosive jet production phase there may
  be steep-spectrum coherent radio bursts which LOFAR would be well
  placed to observe.}
\label{ymaxi}
\end{figure}

These goals and speculations are summarized in Fig \ref{ymaxi}.

To end, within the LOFAR project, and the Transients Key Science
Project (Fender et al. 2008) in particular, we look forward to a
collaboration with MAXI which will be very profitable for the whole
high-energy astrophysics community.

\section*{References}

\re Cabanac C., Fender R.P., Dunn R.J.H., K\"ording E., 2008, MNRAS
submitted

\re Corbel S., Fender R.P., Tzioumis A.K., Tomsick J.A., Orosz J.A.,
Miller J.M., Wijnands R., Kaaret P., 2002, Science, 298, 196

\re Fender R., 2006, In: Compact stellar X-ray sources. Edited by
Walter Lewin \& Michiel van der Klis. Cambridge Astrophysics Series,
No. 39. Cambridge, UK: Cambridge University Press, p. 381--419 {\bf
  (astro-ph/0303339)}

\re
Fender R., Belloni T., Gallo E., 2004, MNRAS, 355, 1105

\re
 Fender R., Wijers R., Stappers B., et al., 2008, In
  "Bursts, Pulses and Flickering: wide-field monitoring of the dynamic
  radio sky", Tzioumis, Lazio \& Fender (Eds), Proceedings of Science
  {\bf (arXiv:0805.4349)}

\re Gallo E., 2007, THE MULTICOLORED LANDSCAPE OF COMPACT OBJECTS AND
THEIR EXPLOSIVE ORIGINS. AIP Conference Proceedings, Volume 924,
pp. 715-722 (2007) {\bf (astro-ph/0702126)}

\re
Gallo E., Fender R., Pooley G.G., 2003, MNRAS, 344, 60

\re
Grimm H.-J., Gilfanov M., Sunyaev R., 2002, MNRAS, 391, 923

\re
Gunst A., van der Schaaf K., Bentum M.J., published in the Proceedings
from SPS-DARTS 2006, The second annual IEEE BENELUX/DSP Valley Signal
Processing Symposium, March 28-29 (2006), Antwerp, Belgium

\re
Homan J., Belloni T., 2005, Ap\&SS, 300, 107

\re
K\"ording E., Falcke H., Corbel S., 2006, A\&A, 456, 439

\re
K\"ording E., Fender R., Migliari S., 2006, MNRAS, 369, 1451

\re
K\"ording E., Jester S., Fender R., 2006, MNRAS, 371, 1366

\re
Kawai N., et al., 1999, Astronomische Nachrichten, vol. 320, no. 4, p. 372

\re Lorimer D., Bailes M., McLaughlin M.A., Narkevic D.J., Crawford
F., 2007, Science, 318, 777

\re McHardy I.M., K\"ording E., Knigge C., Uttley P., Fender R.P.,
2006, Nature, 444, 730

\re
Matsuoka M. et al., 2007, SPIE, 6686, 32

\re
Merloni A., Heinz S., di Matteo T., 2003, MNRAS, 345, 1057

\re
Mirabel I.F., Rodriguez L.F., Cordier B., Paul J., Lebrun F., 1992, Nature, 358, 215

\label{last}

\end{document}